# Pill Identification using a Mobile Phone App for Assessing Medication Adherence and Post-Market Drug Surveillance

by   David M. Prokop [a,b],    (Correspondence Author)

Joseph B. Babigumira [c],

Ashleigh Lewis [d]

[a] PI, TruMedicines Ai, Sammamish, WA., United States
[b] BSc. School of Health & Computer Science, University of Waterloo, Waterloo, ON, Canada
[c] PI, Assoc Prof, Global Health, Adjunct Assoc Prof, Pharmacy, University of Washington, Seattle,WA,USA
[d] REDCap Administrator, BA University of Washington, ITHS Medicine Research IT, Seattle, WA, USA

# Pill Identification using a Mobile Phone App for Assessing Medication Adherence and Post-Market Drug Surveillance

by David M. Prokop [a,b], Joseph B. Babigumira [c], Ashleigh Lewis [d]

[a] PI, TruMedicines Ai, Sammamish, WA., United States

[b] BSc. School of Health & Computer Science, University of Waterloo, Waterloo, ON, Canada

[c] PI, Assoc Prof, Global Health, Adjunct Assoc Prof, Pharmacy, University of Washington, Seattle,WA,USA

[d] REDCap Administrator, BA University of Washington, ITHS Medicine Research IT, Seattle, WA, USA

**ABSTRACT**

**Objectives:** Medication non-adherence is an important factor in clinical practice and research methodology. There have been many methods of measuring adherence yet no recognized standard for adherence. Here we conduct a software study of the usefulness and efficacy of a mobile phone app to measure medication adherence using photographs taken by a phone app of medications and self-reported health measures.

**Method:** A total of 34 participants were recruited to participate in the study. Out of these, 26 (76.5%) participants were able to complete study questionnaire while 8 (23.5%) were not able. Participants volunteered in a study of the efficacy of a mobile phone app for medication adherence and measurement of mood and health measures. Participants used their own android mobile phones and downloaded our phone application. Participants were instructed to take a photo of 1 harmless placebo pill (breath mints) a day for 2 weeks, and record their answers linked to a Research Electronic Data Capture (REDCap™)[1] data collection server using MyCap™[2] enabled phone application. Pill image and self-reported health measurements were recorded daily. Participants were paid for a stipend for participation.

**Results:** The participants were asked by the app 'would help to keep track of your medication', their response indicated 92.9% felt the app 'would you use this app every day' to improve their medication adherence. The subjects were also asked by the app if they 'would photograph their pills on a daily basis'. Subject responses indicated 63% would use the app on a daily basis. By using the data collected, we determined that subjects who used the app on daily basis were more likely to adhere to the prescribed regimen.

**Conclusions:** Pill photographs are a useful measure of adherence, allowing more accurate time measures and more frequent adherence assessment. Given the ubiquity of mobile telephone use, and the relative ease of this adherence measurement method, we believe it is a useful and cost-effective approach. However we feel the 'manual' nature of using the phone for taking a photograph of a pill has individual variability and an 'automatic' method is needed to reduce data inconsistency.

## 1. Introduction

Medication non-adherence is a common and serious problem that reduces therapeutic efficacy and increases healthcare costs and contributes to deaths. In clinical trials, non-adherence may lead to erroneously accepting or rejecting claims of efficacy. Accurate measurement of adherence is, therefore, important, and although many methods have been used to assess adherence, all have drawbacks. Patients and research subjects may be reluctant to admit to non-adherence and self-reporting data overestimate adherence by as much as 300% [4]. Treating physicians' estimates of adherence are also poor. One study found that healthcare professionals overestimate adherence by 50% [5]. Traditional counts of dosage units, for example, capsules, do not provide information on when capsules are taken, or even if they were taken at all [6].

## 2. Method

### 2.1. Software

Research Electronic Data Capture (REDCap™) is a browser-based, metadata-driven software and workflow methodology for designing clinical and translational research databases.[1] It is widely used in the academic research community: the REDCap™ Consortium is a collaborative, international network of more than 2400 institutional partners in over 115 countries, with more than 590,000 total end-users employing the software for more than 450,000 ongoing research studies.[2]

MyCap™ is a mobile phone application developed by staff at Vanderbilt University. MyCap™ is a iPhone® and Android application that can be downloaded from the Apple Store® and Google Play®. It leverages Apple's Research Kit® (ResearchStack™ on Android) and the device's built-in sensors to collect participant information directly from study participants. MyCap™ leverages REDCap™, ResearchKit™, and ResearchStack™ to capture participant/patient reported outcomes via mobile devices. REDCap™ is used to define tasks/instruments/surveys to be completed by participants. MyCap™ translates REDCap™ task metadata into a structure compatible with ResearchKit™ and ResearchStack. When a project participant completes a task, MyCap™ converts the results into a format compatible with REDCap™ before synchronizing back to the REDCap™ project.[3]

### 2.2. Study Setting and Participants

The study administration took place in at TruMedicines Company (Sammamish, WA) and the University of Washington, (Seattle, WA) from October 2018 to September 2019. Study participants were randomly recruited from a lounge/dining area at the University of Washington Health Sciences facility. A total of 34 participants were recruited to participate in the study. Out of these, 26 (76.5%) participants were able to complete study questionnaire while 8 (23.5%) were not able. (14 male, 12 female) volunteered to participate. Each participant were asked to take part in a 2 week trial of a MyCap™ mobile phone application created to track medications and health measures. Each participant was given a box of 14 harmless placebo medications (breath mints), individually packaged "pills" in pill packets. Participants were instructed to take

1 pill per day (7 capsules per week). Participants took placebo medication over 2 weeks. Participants were told they would earn a $20 gift certificate at the end of the study period. This study was approved by the University of Washington Institutional Review Board (IRB) as a Non-Significant Risk (NSR) study.

Participants used their own android phones and were provided a labeled box of placebo medications with a Quick Response (QR) code to allow them to register into the study and input their health measure answers into the REDCap™ study database. The mobile phones were not modified and served as fully functional voice and text communication systems, as participants were able to make personal phone calls and send text messages to study administrators. Participants were given a $25 dollar gift card if they successfully completed the 2 weeks study period.

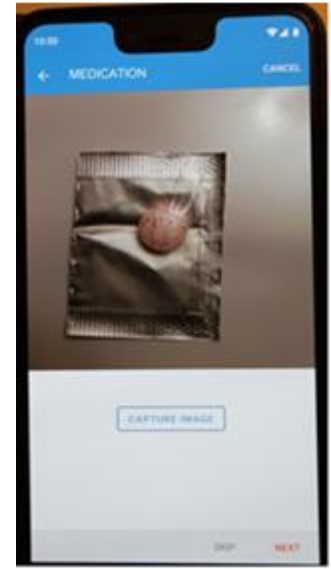

Photo #1 Pill Image

Participants were instructed to photograph their placebo medication "dose" each day (Photo #1) and answer the accompanying health related questions. They were instructed to pull one pill packet from their pill box, remove the pill packet from the box and photograph the placebo pill before consuming the pill. Each image embedded a time stamp indicating when the medication was held by the study participant.

## 3. Results

Results in Table1 show that majority (73.1%) of the participants were aged between 18 and 34 years old. Those who were aged between 35 and 54 years old comprised of 23.1% and only one participant was aged above 55 years. Results have indicated that more than half of the participants (53.9%) were males and only 46.2% were females.

Findings have further indicated that 12 (52.2%) of the respondents always took their medications during the morning hours, 6 (26.1%) took their medication both in the morning and night while only 5(21.7%) always took their medication dose during the night hours.

The results have further indicated that the average stress level for the participants is 4.42 (2.23). The stress level was further categorized three levels; (1) none (zero stress level, (2)-between 1-4 grouped as low stress level and (3) between 5-8 grouped as high stress level. The table has shown that 15 (57.7%) of the participants were experiencing high stress level, 10 (38.5%) of the respondents were experiencing low/moderate stress level and only one participants was not stressed during the time of the study (was experiencing zero levels of stress).

Table 1: Socio-demographic characteristics of the participants

| Characteristics | Frequency (N) | Percent (%) |
| --- | --- | --- |
| **Age** | | |
| 18-34 | 19 | 73.1 |
| 35-54 | 6 | 23.1 |
| 55+ | 1 | 3.9 |
| **Gender** | | |
| Male | 14 | 53.9 |
| Female | 12 | 46.2 |
| **Medication time** | | |
| In morning | 12 | 52.2 |
| Both morning &night | 6 | 26.1 |
| At night | 5 | 21.7 |
| | | |
| **Stress, mean(SD)** | 4.42 | 2.23 |
| **Stress level** | | |
| None | 1 | 3.9 |
| low | 10 | 38.5 |
| high | 15 | 57.7 |

Fig1 shows the number of participants who used the recap repeat instrument and their corresponding average number of repeat instances. Participants used more than one instrument. The results have revealed that all the participants who completed the questionnaire used the introduction survey instrument and it was averagely used once by each individual. Out of 26 participants who were involved in the study, 21 (80.77) used the medication instrument and the mean time each individual used this instrument was 9 times. The findings have further revealed that 23 (88.5%) used the Daily task instrument and the average number of times each individual used this instrument was 9 times. Only one participant used the monthly questionnaire instrument. The results in fig 1 further indicated that 11(42.3%) used the app appropriateness tool and each individual used this instrument approximately once.

**Fig 1: Frequency and Number of participants per REDcap™ repeat instrument**

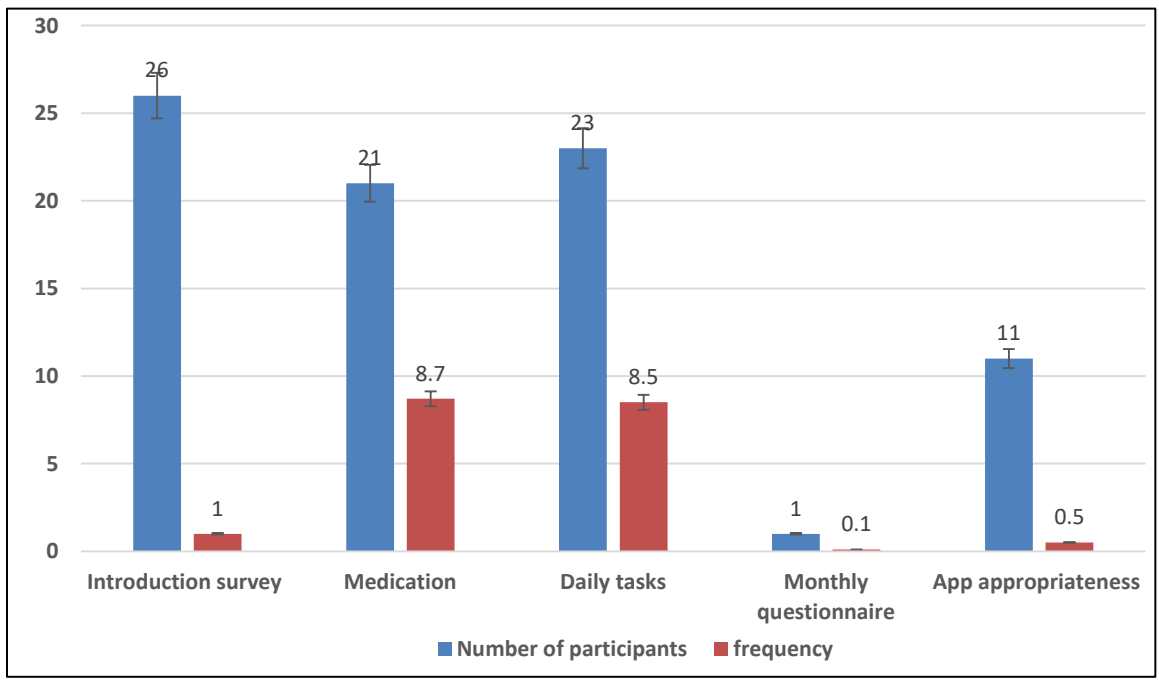

Table 2 shows the relationship between respondent`s choice of the recap read instrument and different socio demographic factors. Results have revealed that the choice of REDCap™ read instrument was not dependent on most of the demographic factors for instance gender, time of medication and respondent`s stress level. However, only age had a statistically significant relationship with choice of the monthly questionnaire as a REDcap™ read instrument. Only one individual aged above 55 years used monthly questionnaire as a REDcap™ read instrument.

**Table2: Relationship between recap read instrument and demographic characteristics**

| Characteristic | Medications | | Daily tasks | | Monthly questionnaire | | App appropriateness | |
|---|---|---|---|---|---|---|---|---|
| | Yes | No | Yes | No | Yes | No | Yes | No |
| **Age** | | | | | | | | |
| 18-34 | 16 (84.2) | 03 (15.8) | 17 (89.5) | 02 (10.5) | 0 (0.0) | 19 (100.0) | 07 (36.8) | 12 (63.2) |
| 35-54 | 04 (66.47) | 02 (33.3) | 05 (83.3) | 01 (16.7) | 0 (0.0) | 6 (100.0) | 03 (50.0) | 03 (50.0) |
| 55+ | 01 (100.0) | 0 (0.00) | 1 (100.0) | 0 (0.00) | 1 (100.0) | 0 (0.0) | 01 (100.0) | 00 (0.0) |
| | P value =0.652 | | P value = 0.859 | | P value = 0.000** | | P value=0.419 | |
| **Gender** | | | | | | | | |
| Male | 12 (85.7) | 02 (14.3) | 12 (85.7) | 02 (14.3) | 1 (7.1) | 13 (92.9) | 07 (50.0) | 07 (50.0) |
| Female | 09 (75.0) | 03 (25.0) | 11 (91.7) | 01 (8.3) | 0 (0.0) | 12 (100.0) | 04 (33.3) | 08 (66.7) |
| | P value = 0.490 | | P value = 0.636 | | P value = 0.345 | | P value=0.391 | |
| **Medication time** | | | | | | | | |
| Morning | 09 (81.8) | 02 (18.2) | 10 (90.9) | 01(90.9) | 0 (0.0) | 1 (100.0) | 1 (100.0) | 0 (0.0) |
| Morning & night | 04 (66.7) | 02 (33.3) | 04 (66.7) | 02 (33.3) | 01 (10.0) | 9 (90.0) | 5 (50.0) | 05 (50.0) |
| At night | 05 (100.0) | 0 (0.00) | 05 (100.0) | 0 (0.0) | 15 (100.0) | 15 (100.0) | 05 (33.3) | 10 (66.7) |
| | P value = 0.361 | | P value = 0.228 | | P value = 0.435 | | P value = 0.350 | |
| **Stress level** | | | | | | | | |
| None | 1 (100.0) | 0 (0.0) | 1 (100.0) | 0 (0.0) | 01 (9.09) | 10 (90.9) | 06 (54.6) | 05 (45.52) |
| Low | 7 (70.0) | 3 (30.0) | 10 (100.0) | 0 (0.0) | 0 (0.0) | 6 (100.0) | 02 (33.3) | 04 (66.7) |
| High | 13 (86.7) | 02 (13.3) | 12 (80.0) | 3 (20.0) | 0 (0.0) | 5 (100.0) | 02 (40.0) | 03 (60.0) |
| | P value = 0.517 | | P value = 0.288 | | P value =0.592 | | P value=0.676 | |

The study further explored the different side effects that were experienced by the individuals when taking different kind of medicine. Table 3 shows the number of times of occurrence of each side effect. Majority of the individuals, 24 (92.3%) didn't experience any side effects for instance nausea, vomiting, diarrhea and heart burn when taking TruMedicines. Findings form table 3 have indicated that only 1 (3.9%) experienced nausea as side effects. Also 1 (3.9%) experienced nausea as a side effect 6 times when taking TruMedicines. Only 2(7.7%) of the respondent experienced vomiting once when taking the medicine. Results have shown that only 1(3.9%) experienced diarrhea once when taking the TruMedicines while 1 (3.9%) experienced diarrhea four times when taking medicine. Only 2 (7.7%) of the respondent who were taking the TruMedicines experienced heart burn as a side effect once. Findings have also shown that only 1(3.9%) individuals experienced other side effects rather that those mentioned in the questionnaire.

**Table3: Occurrence of Side Effects**

| Side effects | No of times, N (%) | | | |
|---|---|---|---|---|
| | 0 | 1 | 4 | 6 |
| **Nausea** | 24 ( 92.3 ) | 1 (3.9) | 0 (0.0) | 1 (3.9) |
| **Vomiting** | 24 ( 92.3 ) | 2 (7.7) | 0 (0.0) | 0 (0.0) |
| **Diarrhea** | 24 ( 92.3 ) | 1 (3.9) | 1 (3.9) | 0 (0.0) |
| **Heart burn** | 24 ( 92.3 ) | 2 (7.7) | 0 (0.0) | 0 (0.0) |
| **Others** | 25 (96.2) | 0 (0.0) | 1 ( 3.9) | 0 (0.0) |

The study also explored the daily tasks that individuals do on a day to day basis as shown in table 4. Results have indicated that on average, each individual spends 8 hours of sleep each night and spends 30 minutes exercising their bodies every day. Also on average, each individual drinks 26 ounces of water daily and the mean stress level for each individual was 5 with a range of 1-8

**Table 4: Descriptive Statistics of Daily tasks**

| Daily tasks | Observations | Mean (SD) | Max | Min |
|---|---|---|---|---|
| Daily sleep hours | 24 | 7.6 (1.01) | 10 | 6 |
| Daily exercise minutes | 23 | 30.0 (20.8) | 79 | 0 |
| Daily ounces of water | 21 | 26.2 (18.7) | 64 | 0 |
| Stress level | 23 | 4.5 (1.9) | 8 | 1 |

Table 5 shows the association between different daily tasks and the socio demographic factors. Gender and time of taking medication were statistically significant with the number of minutes of daily exercise. Result show that on average, females spend more 18 minutes exercising daily than the males. Also, those individuals who take their medication at night spend more hours exercising as compared to those who take their medication in the morning.

**Table 5: Relationship between daily tasks and demographic factors**

| Characteristic | Hours of daily sleep, mean (SD) | Minutes of daily exercise, mean (SD) | Ounces of daily water, mean (SD) | Stress level, mean (SD) |
|---|---|---|---|---|
| **Age** | | | | |
| 18-34 | 7.8 (1.1) | 26.8 (20.9) | 25.4 (19.4) | 4.76 (1.98) |
| 35-54 | 7.2 (0.8) | 42.8 (19.4) | 28.5 (20.8) | 3.6 (1.8) |
| 55+ | 7.0 (0.0) | 23.0 (0.0) | 30 (0.0) | 5.0 (0.0) |
| | P value =0.454 | P value = 0.856 | P value = 0.944 | P value=0.499 |
| **Gender** | | | | |
| Male | 7.7 (1.1) | 20.8 (4.0) | 24.7 (5.6) | 4.8 (1.7) |
| Female | 7.6 (0.99) | 38.6 (6.7) | 27.6 (6.2) | 4.3 (2.10 |
| | P value = 0.846 | P value = 0.04** | P value = 0.635 | P value=0.493 |
| **Medication time** | | | | |
| In the morning | 7.7 (1.2) | 22.2 (15.0) | 19.5 (14.6) | 4.2 (20.0) |
| Morning & night | 7.4 (0.6) | 23.6 (7.3) | 37.0 (20.4) | 4.4 (1.8) |
| At night | 7.6 (0.6) | 55.8 (17.8) | 45.3 (16.7) | 5.3 (2.6) |
| | P value = 0.840 | P value = 0.003** | P value = 0.050 | P value = 0.350 |

Respondents also answered some questions about the app usability as shown in table 6. Results have shown that almost all, 09 (90.0%) found the app easy to use and majority 3 (75.0%) of these were males. More than half, 06 (60.0%) of the respondents responded that they would use the app every day if they were taking medication in the future. Majority of the respondents (80.0%) found direct pill scanning useful to them. Findings have further indicated that most of the respondents (80.0%) reported that the medication progress charts helped them to stay on track when taking the placebo pills. Also majority 09 (90.0%) of the respondents felt like they were in control of their medication therapy.

Table6: Usability of the App

| usability questions | Overall, N (%) | Male, N (%) | Female, N (%) |
|---|---|---|---|
| App easy | 09 (90.0) | 06 (100.0) | 3 (75.0) |
| App everyday | 06 (60.0) | 4 (66.7) | 2 (50.0) |
| App scanning pill | 8 (80.0) | 6 (100.0) | 2 (50.0) |
| App charts | 8 (80.0) | 5 (83.3) | 3 (75.0) |
| Feeling in control | 9 (9.0) | 6 (100.0) | 3 (75.0) |
| App medication | 9 (90.0) | 6 (100.0) | 3 (75.0) |

Of the 26 participants, 14 (51%) were men and 12 (49%) were women. Participant's age range was <18yrs (0.0%), 18-34yrs (19, 70.4%), 35-54 (7, 25.9%), 55+ (1, 3.7%). Participants owned a mobile phone and were familiar with modern mobile phone applications. Participants were able to use their phone to scan the study access QR code and install the application without assistance. Overall adherence percentage of prescribed placebo medication tracked estimated by self-reported daily count was 92.9%. (Chart #1)

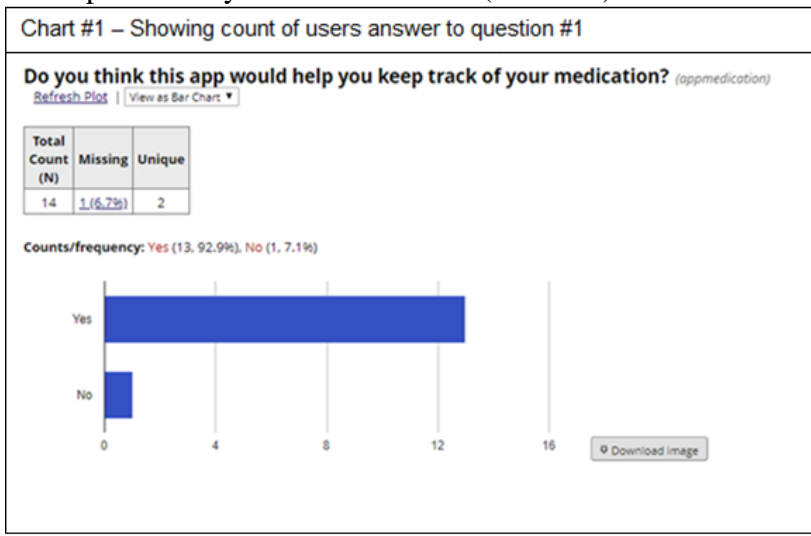

Participants self-reported feeling in 'control' of their medication as shown by a (92.9% , Yes count in Chart #2.

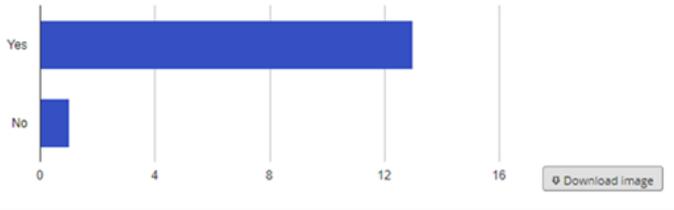

Chart # 2 - Showing Counts of users answers to question # 2

Participants self-reported a feedback of a daily historical chart of their medication progress helped them improve their medication adherence or as stated 'stay on track'. As shown by a 85% self-reported yes count in Chart #3.

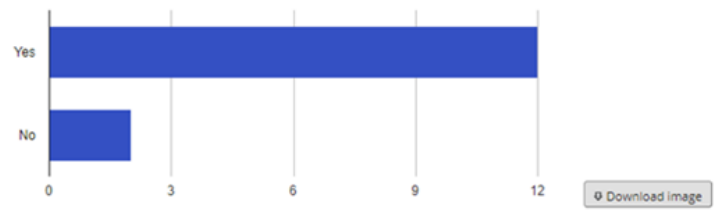

Chart # 3 - Showing counts of users answers to question # 3

## 4. Discussion

It is important to highlight that the current mobile health app market is generally unregulated by any government or reporting agencies. In general, phone application 'online stores' provide guidelines about restricted content, privacy, and security of the data, and monetization of the apps; however, these guidelines are not a quality control assessment of the available apps. Recently, the US Food and Drug Administration (FDA) released a guidance document stating which type of mobile medical apps will be subject to their regulation [7]. However, at this stage, apps to promote medication adherence will not be within the FDA regulation oversight, as these apps are not intended to provide diagnosis or treatment recommendations.[8]

"For short-term treatments several quite simple interventions increased adherence and improved patient outcomes, but the effects were inconsistent from study to study with less than half of studies showing benefits. Current methods of improving adherence for chronic health problems are mostly complex and not very effective, so that the full benefits of treatment cannot be realized. High priority should be given to fundamental and applied research concerning innovations to assist patients to follow medication prescriptions for long-term medical disorders".[9]

The ability of mobile telephones to save and transmit photographs as they are collected allows for more frequent assessment of adherence. Use of mobile phones for clinical purposes may be seen as an analog of industry's successful development of other technology to monitor medication use and effects, for example, take-home devices for measurement of blood pressure, pulmonary function, glycemic index, and anticoagulant activity. In addition to adherence tracking, photographs with accurate timestamps of medication provide more accurate tracking by online database of at-home medication use on measurable, at-home self-reported physiological effects. Other capabilities of mobile phones invite investigation for clinical use; for example, image recognition of pill identify or type, tele medication video calls of at-home treatment and symptoms, text messaging for symptom severity tracking, and patient reminders.  Trials without incentives might show different adherence patterns. Incentives have been shown to increase adherence whether measured by self-report, pill count, or electronic monitoring [10].

Mobile phones themselves might provide a convenient vehicle for incentives for desired behaviors, for example, credits to mobile phone billing or other micropayments, free downloads of phone applications, increased mobile phone time, or text messaging limits, etc. Study limitations include our small sample size, a larger trial would be better suited to detect meaningful differences and similarities between adherence measures. Finally, administration of a marker of adherence that can be detected in the urine [11] would permit a more precise estimate of the utility of this technique. For patients who already own camera-equipped mobile telephones, instructing them to e-mail photographs is an inexpensive and easily implemented method of measuring medication adherence. Innovative uses of mobile telephones offer researchers and clinicians new ways to improve clinical trials and practice. Pill photographs are a useful measure of adherence, allowing more accurate time measures and more frequent adherence assessment. Given the ubiquity of mobile telephone use, and the relative ease of this adherence measurement method, we believe it is a useful and cost-effective approach. However we feel the 'manual' nature of using the phone for taking a photograph of a pill has individual variability and an 'automatic' method is needed to reduce medication image inconsistency

## 5. Conclusions

Pill photographs are a useful measure of adherence, allowing more accurate time measures and more frequent adherence assessment. Given the ubiquity of mobile telephone use, and the relative ease of this adherence measurement method, we believe it is a useful and cost-effective approach. However we feel the 'manual' nature of using the phone for taking a photograph of a pill has individual variability and an automatic method is needed to reduce medication image inconsistency.

.
**Acknowledgements**
*The authors thank the generous support of the National Institute of Health the FDA Grant 1R43FD006302-01, the staff of the University of Washington ITHS Medical Research Dept, our research participants, and this article's anonymous reviewers for their helpful comments and suggestions.*

**Disclosure Statement**
*The author(s) have no conflicts of interest to disclose.*